# Confidentiality without Encryption For Cloud Computational Privacy

Chaffing and Winnowing in Computational-Infrastructure-as-Service

*Sashank Dara*
Research Scholar, IIIT - Bangalore
Bangalore, India
krishna.sashank@gmail.com

*Abstract—* Advances in technology has given rise to new computing models where any individual/organization (Cloud Service Consumers here by denoted as CSC's) can outsource their computational intensive tasks on their data to a remote Cloud Service Provider (CSP) for many advantages like lower costs , scalability  etc. But such advantages come for a bigger cost "Security and Privacy of data" for this very reason many CSC's are skeptical to move towards cloud computing models.

While the advances in cryptography research are promising, there are no practical solutions yet for performing any operations on encrypted data [1]. For this very reason there is strong need for finding alternative viable solutions  for us to benefit from Cloud Computing.

A technique to provide confidentiality without encryption was proposed in the past namely "Chaffing and Winnowing: Confidentiality without Encryption" by Ronald L. Rivest [2]. While this technique has been proposed for packet based communication system, its not adaptable in all cloud service models like Software-as-Service, Platform-as-Service or Infrastructure-as-Service [3].

In this paper we propose an adaptation of this technique in a cloud computational setup where CSC's outsource computational intensive tasks like web log parsing, DNA Sequencing etc to a MapReduce like CSP service.

## I. INTRODUCTION

### A. Computational Intensive Tasks that can be OutSourced

Computational intensive tasks can be outsourced to a remote cloud service provider for leveraging the umpteen benefits that cloud computing promises like scalability, afford ability etc. Out of many such tasks that can be outsourced we consider few types of tasks that fit in the model of  MapReduce framework.

In the Internet advertising business, like whether it is on line shopping website, social networking website or a simple real estate search site , the customer event logs are critical to understanding customer behavior patterns, such as duration of their session , the frequency and amount of time spent on different pages and trending search terms. Many digital advertising and marketing firms segment users and customers based on the collection and analysis of data from browsing session logs.  Parsing such logs is very computational intensive often running into Tera bytes of data and the logs are very crucial  often containing sensitive information of the customers. So outsourcing such task to a CSP is very risky as such information can fall in wrong hands either through external attacks on CSPs or internal malicious people.

We emphasize that our paper is constructed throughout in the context of Weblog parsing, although it can be adopted in similar *Computational-Infrastructure-as-Service* (here after referred as Computational-IaaS) like services without loss of generality. A verbose list of other use cases is given here[5]

### B. MapReduce like  Computational-IaaS

MapReduce[4] is a framework for processing highly distributable problems across huge datasets using a large number of computers. Users specify a map function that processes a key/value pair to generate a set of intermediate key/value pairs, and a reduce function that merges all intermediate values associated with the same intermediate key. Use cases for such a framework would be log parsing, image processing, (re)calculating indexes of huge corpus of documents like web, scientific computing etc[5].

Such computational infrastructure is being offered as service by Amazon recently as Elastic MapReduce[6] along with their cloud services.  Many case studies on the same are featured on their  website[7]. Such service can be offered by any CSP.

A typical architecture of a  such CSC, especially for web log parsing, would have few major parts, as shown in Figure 1.

1. **Data Collection**: A large scale data collecting, aggregating  and data moving  service such as Apache Flume [8]  or  Facebook's Scribe [9] or Chukwa [10] or a home brewn tools. Such service typically has three tiers, the agent tier which is responsible for collecting data from various sources internal to CSC, the collecting tier which is responsible for aggregating such data, the streaming tier which streamlines the data to a CSP.
2. **Data Parsing**: A Hadoop like MapReduce service

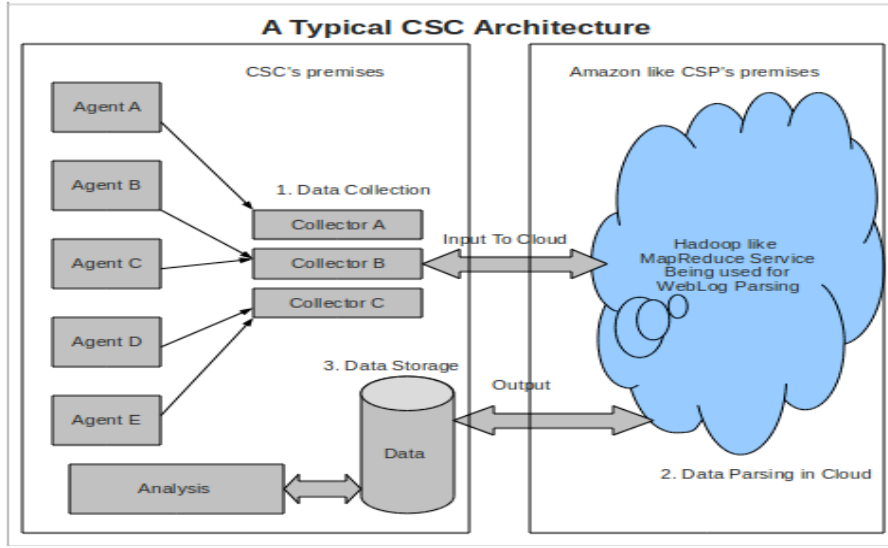

Figure 1 : Overview of Typical CSC Architecture

capable of running on cheap commodity hardware for carrying out computational intensive task in a distributed file system. The MapReduce is a programmable framework that gives high flexibility of executing a particular task say parsinga log files of various formats.

3. **Data Storage**: The output of such a data parsing service can be exported into a CSC specific Storage.
4. **Data Analysis:** Finally the data is analysed from the stored database using Apache Hive , Apache Pig or SQL like Query languages.

A CSC can outsource the computational intensive data parsing work to a CSP to leverage the benefits of cloud computing, the data aggregation is done within the CSC's premises and the final data storage can be outsourced or can be retained within CSC's premises. A CSC who is conscious of security and privacy of the data can retain the data storage back within its premises without outsourcing it to a CSP. Our technique works in such setup where CSC retains the data storage resulting from CSP's computational task.

C. *Chaffing and Winnowing*

The term derived from agriculture is used to denoted the process of separating wheat from chaff. As stated earlier this technique is first proposed for achieving confidentiality without encryption when sending data over an insecure channel.

Briefly the two parties communicating share a common secret key priorly , the sender sends the actual packets signed with shared secret along with few other packets with some arbitrary value as signature, the receiver then separates the good packets from bad by calculating the signature of each packet with pre shared key and comparing it with the received value. Thus a passive on looker cannot make out good packets from bad packets since they do not have the shared secret. Here technically the packets are not encrypted. They are in clear text but obfuscated along with some other fake packets.

We show that such a technique can be adapted for achieving cloud computational privacy particularly in the context of Computational-Infrastructure-as-service because operating on encrypted data is not yet feasible on a remote CSP server.

II. OVERVIEW

At an outset, CSC obfuscates the real data with fake data after appropriately tagging it before sending it to a CSP, the CSP agnostic of the fact of such obfuscation parses religiously the log files, the CSC later separates the real results from fake results using the tags. Any adversary, either an external or an internal attacker of CSP cannot make out the real data from fake data due to lack of the appropriate tags. We shall detail how it is done in next section.

III. ARCHITECTURE

Below are the changes to be done at each phase in the entire cycle to achieve confidentiality of CSC's data without actually encrypting it.

1. Data Collecting: CSC has multiple agents for data collecting the data. All such agents can share a common secret key and such key can be used to sign each of the log file generated by them. CSC also can have few fake agents that generate fake data and sign the data with fake key. All such log files are aggregated by collector nodes and streamlined to be sent to MapReduce Cluster.
2. Data Parsing: The MapReduce Cluster parses the weblogs being agnostic of such fake data among the real data and returns the results to CSC's data Storage. Here care has to be taken while programming the MapReduce framework to tag the appropriate signatures to real and fake data.
3. Data Storage: CSP's output is then exported to CSC

specified data storage for later analysis.

4. Data Analysing: CSC's analyser shares the same secret key as the real agents. CSC's analyser carefully ignores from the results the ones generated by fake data by calculating the signature.

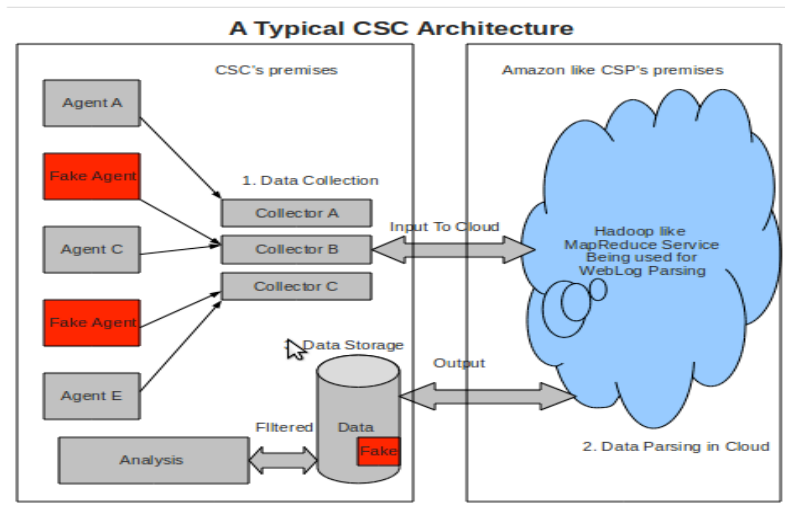

Figure 2: A CSC Architecture with Fake Agents

Any adversary, an external attacker of CSP or an untrusted CSP itself cannot make out the real data from fake data due to lack of the secret key.

It can be noted that, the level of confidentiality required by the CSC can be achieved by increasing the number of fake agents. Also intuitively, the log parsing is an intensive task but the not the fake log generation.

## IV. CONCLUSIONS AND FURTHER WORK

With little additional expense to be spent on parsing the fake logs, we can achieve moderate levels of privacy. Higher degrees of privacy can be achieved too by increasing the number of fake agents without the budget going overboard.

It has to be analysed further on how this technique can be adapted in other Cloud Service Models like Software-as-Service, Platform-as-Service and other Infrastructure-as-Service offerings of CSP.

## ACKNOWLEDGMENT

Sincere thanks to my colleague Saugat Mazumdar and my thesis advisor Dr Muralidhara V.N for initial discussions on this paper.

## REFERENCES

[1] Vinod Vaikuntanathan, New Developments in Fully Homomorphic Encryption,FOCS 2011
[2] Ronald L.Rivest ,"Chaffing and Winnowing:Confidentiality without Encryption", http://theory.lcs.mit.edu/~rivest/chaffing.txt
[3] Tim Mather, Subra Kumaraswamy, and Shahed Latif ,"Cloud security and privacy", Orielly,2009, pp 16
[4] Jean Dean and Sanjay Ghemawat,"MapReduce: Simplified Data Processing on Large Clusters", OSDI'04: Sixth Symposium on Operating System Design and Implementation, San Francisco, CA, December, 2004
[5] Apache Hadoop Users list http://wiki.apache.org/hadoop/PoweredBy
[6] Amazon Elastic MapReduce http://aws.amazon.com/elasticmapreduce/
[7] Common Hadoopable Problems http://www.slideshare.net/cloudera/20100806-cloudera-10-hadoopable-problems-webinar-4931616
[8] Apache Flume https://cwiki.apache.org/FLUME/
[9] Facebook's Scribe https://github.com/facebook/scribe/wiki/
[10] Hadoop Chukwa http://wiki.apache.org/hadoop/Chukwa